\documentclass[twocolumn,showpacs,preprintnumbers,amsmath,amssymb]{revtex4}

\usepackage{graphicx}
\usepackage{dcolumn}
\usepackage{bm}

\begin{document}

\title{Recent $\alpha$ decay half-lives and analytic expression predictions including \\
the superheavy nuclei}

\author{G. Royer$^{1}$}
\email{royer@subatech.in2p3.fr}
\author{H.F. Zhang$^{2}$}

\affiliation{\footnotesize $^1$ Laboratoire Subatech, UMR:
IN2P3/CNRS-Universit\'e-Ecole des Mines, 4 rue A. Kastler,
 44307 Nantes Cedex 03, France}

\affiliation{\footnotesize $^2$ School of Nuclear Science and
Technology, Lanzhou University, Lanzhou 730000, People's Republic of China}

\date{\today}

\begin{abstract}
New recent experimental $\alpha$-decay half-lives have been compared with the results obtained from previously  proposed formulae depending only on the mass and charge numbers of the $\alpha$ emitter and the $Q_{\alpha}$ value. For the heaviest nuclei they are also compared with calculations
using the Density-Dependent M3Y (DDM3Y) effective interaction and
the Viola-Seaborg-Sobiczewski (VSS) formulae. The correct agreement allows to provide predictions for the $\alpha$ decay half-lives of other still unknown superheavy nuclei from these analytic formulae using the extrapolated $Q_{\alpha}$ of Audi, Wapstra and Thibault \cite{Au03}. 
\end{abstract}

\maketitle

\begin{table*}[h]
\label{table1} \caption{Comparison between recently known experimental $\alpha$
decay half-lives and results obtained
 with the formulae (1-4).}
\begin{ruledtabular}
\begin{tabular}{lllllllll}
Nucleus& Q$^{expt}_{\alpha}$(MeV)& T $^{expt}_{1/2,\alpha}$(s)& T $^{Formulae}_{1/2,\alpha}$(s)&Nucleus&  Q$^{expt}_{\alpha}$(MeV)& T $^{expt}_{1/2,\alpha}$(s)& T $^{Formulae}_{1/2,\alpha}$(s)   \\
\hline
$^{105}$Te&4.900(0.050)&0.70$^{+0.25}_{-0.17}\times10^{-6}$&0.37$^{+0.21}_{-0.13}\times10^{-6}$&$^{156}$Er&3.486&2.3$\times10^{10}$&1.6$\times10^{10}$ \\
$^{158}$Yb&4.172&4.3$\times10^{6}$& 2.7$\times10^{6}$&$^{160}$Hf&4.902&1.9$\times10^{3}$&1.9$\times10^{3}$ \\
$^{174}$Hf&2.497&6.3$\times10^{22}$& 4.55$\times10^{23}$& $^{158}$W&6.612$^{+0.003}_{-0.003}$& 1.5$^{+2}_{-2}\times10^{-3}$& 1.21$^{+0.03}_{-0.03}\times10^{-3}$ \\
$^{168}$W&4.507&1.6$\times10^{6}$&3.1$\times10^{6}$&$^{162}$Os&6.767$^{+0.003}_{-0.003}$& 1.9$^{+2}_{-2}\times10^{-3}$& 2.33$^{+0.06}_{-0.05}\times10^{-3}$ \\
$^{164}$Os&6.475&4.2$\times10^{-2}$&2.2$\times10^{-2}$&$^{166}$Pt&7.286&3.0$\times10^{-4}$&2.85$\times10^{-4}$ \\
$^{168}$Pt&6.997&2.0$\times10^{-3}$&2.2$\times10^{-3}$&$^{170}$Pt&6.708&1.4$\times10^{-2}$&2.0$\times10^{-2}$ \\
$^{172}$Hg&7.525&4.2$\times10^{-4}$&2.7$\times10^{-4}$&$^{174}$Hg&7.233&2.1$\times10^{-3}$&2.0$\times10^{-3}$ \\
$^{188}$Hg&4.705&5.3$\times10^{8}$&2.0$\times10^{8}$&$^{178}$Pb&7.790&2.3$\times10^{-4}$&2.1$\times10^{-4}$ \\
$^{180}$Pb&7.415&5.0$\times10^{-3}$&2.75$\times10^{-3}$&$^{184}$Pb&6.774&6.1$\times10^{-1}$&3.6$\times10^{-1}$ \\
$^{186}$Pb&6.470&1.2$\times10^{1}$&4.7$\times10^{0}$&$^{188}$Pb&6.109&2.7$\times10^{2}$&1.3$\times10^{2}$ \\
$^{190}$Pb&5.697&1.8$\times10^{4}$&8.7$\times10^{3}$&$^{192}$Pb&5.221&3.6$\times10^{6}$&2.1$\times10^{6}$ \\
$^{194}$Pb&4.738&9.8$\times10^{9}$&1.3$\times10^{9}$ &$^{188}$Po&8.087$^{+0.025}_{-0.025}$&4.0$^{+2.0}_{-1.5}\times10^{-4}$&1.1$^{+0.19}_{-0.17}\times10^{-4}$ \\
$^{189}$Po&7.703$^{+0.020}_{-0.020}$&5.0$^{+1}_{-1}\times10^{-3}$&3.0$^{+0.4}_{-0.4}\times10^{-3}$&
$^{190}$Po&7.693&2.5$\times10^{-3}$&1.5$\times10^{-3}$& \\
 $^{192}$Po&7.319$^{+0.011}_{-0.011}$&2.9$^{+1.5}_{-0.8}\times10^{-2}$&2.2$^{+0.2}_{-0.2}\times10^{-2}$&
$^{210}$Po&5.407&1.2$\times10^{7}$&1.0$\times10^{6}$ \\ 
$^{196}$Rn&7.616$^{+0.009}_{-0.009}$&4.4$^{+1.3}_{-0.9}\times10^{-3}$&1.36$^{+0.09}_{-0.09}\times10^{-2}$ &
$^{198}$Rn&7.349&6.5$\times10^{-2}$&9.56$\times10^{-2}$ \\ $^{202}$Ra&8.020&2.6$\times10^{-3}$&3.63$\times10^{-3}$ 
&$^{204}$Ra&7.636&5.9$\times10^{-2}$&5.5$\times10^{-2}$ \\
$^{210}$Th&8.053&1.7$\times10^{-2}$&1.3$\times10^{-2}$& $^{212}$Th&7.952&3.6$\times10^{-2}$&2.4$\times10^{-2}$ \\
$^{218}$U&8.773$^{+0.009}_{-0.009}$&5.1$^{+1.7}_{-1.0}\times10^{-4}$&4.0$^{+0.2}_{-0.2}\times10^{-4}$ &
$^{220}$U&10.30&6.0$\times10^{-8}$&5.8$\times10^{-8}$\\
$^{224}$U&8.620&7.0$\times10^{-4}$&8.2$\times10^{-4}$&$^{226}$U&7.701&5.0$\times10^{-1}$&5.67$\times10^{-1}$\\
$^{228}$Pu&7.950&2.0$\times10^{-1}$&5.13$\times10^{-1}$&$^{230}$Pu&7.180&1.0$\times10^{2}$&2.71$\times10^{2}$ \\
$^{238}$Cm&6.62&2.3$\times10^{5}$&3.3$\times10^{5}$&$^{258}$No&8.151&1.2$\times10^{2}$&5.4$\times10^{1}$\\
$^{258}$Rf&9.25&9.2$\times10^{-2}$&1.03$\times10^{-1}$&$^{260}$Rf&8.901&1.0$\times10^{0}$&1.0$\times10^{0}$\\
$^{266}$Hs&10.34&2.3$\times10^{-3}$&2.1$\times10^{-3}$&$^{270}$Hs&9.02&2.2$\times10^{1}$&1.03$\times10^{1}$\\
$^{270}$Ds&11.2&1.0$\times10^{-4}$&6.7$\times10^{-4}$ &$^{282}$113&10.63$^{+0.08}_{-0.08}$&7.3$^{+13.4}_{-2.9}\times10^{-2}$& 4.27$^{+2.8}_{-1.7}\times10^{-2}$\\
\end{tabular}
\end{ruledtabular}
\end{table*}

\begin{table*}[h]
\label{table2} \caption{Comparison between recent experimental $\alpha$
decay half-lives and results obtained
 with the DDM3Y effective
interaction \cite{Ba04,Ch07}, the GLDM \cite{Roy00,Zh06}, the formulae (1-4) and the VSS expressions \cite{Vio66,Sobi89}.}
\begin{ruledtabular}
\begin{tabular}{llllllll}
Nucleus& Q$^{expt}_{\alpha}$(MeV)& T $^{expt}_{1/2,\alpha}$& T $^{DDM3Y}_{1/2,\alpha}$&T $^{GLDM}_{1/2,\alpha}$& T $^{Formulae}_{1/2,\alpha}$& T $^{VSS}_{1/2,\alpha}$  \\
\hline
&&&&&&& \\
  $^{294}$118&$11.81\pm0.06$& $1.8^{+75}_{-1.3}$ ms &$0.66^{+0.23}_{-0.18}$ ms &$0.15^{+0.05}_{-0.04}$ ms &$0.39^{+0.15}_{-0.11}$ ms&$0.64^{+0.24}_{-0.18}$ ms    \\
  $^{293}$116&$10.67\pm0.06$& $53^{+62}_{-19}$ ms  &$206^{+90}_{-61}$ ms &$22.81^{+10.22}_{-7.06}$ ms &$308^{+136}_{-93}$ ms   &$1258^{+557}_{-384}$ ms     \\
  $^{292}$116&$10.80\pm0.07$& $18^{+16}_{-6}$ ms   &$39^{+20}_{-13}$ ms  &$10.45^{+5.65}_{-3.45}$ ms   &$27^{+14}_{-9}$ ms  &$49^{+26}_{-16}$ ms    \\
  $^{291}$116&$10.89\pm0.07$&$6.3^{+11.6}_{-2.5}$ ms  &$60.4^{+30.2}_{-20.1}$ ms &$6.35^{+3.15}_{-2.08}$ ms &$89^{+46}_{-30}$ ms  &$336.4^{+173.1}_{-113.4}$ ms  \\
 $^{290}$116&$11.00\pm0.08$& $15^{+26}_{-6}$ ms  &$13.4^{+7.7}_{-5.2}$ ms &$3.47^{+1.99}_{-1.26}$ ms &$8.9^{+5.4}_{-3.3}$ ms  &$15.2^{+9.0}_{-5.6}$ ms  \\
  $^{288}$115  &10.61 (6)&87 $^{+105}_{-30}$ ms    &    409            ms  &94.7$^{+41.9}_{-28.9}$ ms  & $582^{+278}_{-187}$ ms        &997$^{+442}_{-303}$  ms    \\
  $^{287}$115  & 10.74 (9)&   32$^{+155}_{-14}$ ms   &   49  ms              &46.0$^{+33.1}_{-19.1}$ ms  & $53^{+38}_{-22}$ ms        & 207$^{+149}_{-85}$  ms   \\
  $^{289}$114& $9.96\pm0.06$& $2.7^{+1.4}_{-0.7}$ s  &$3.8^{+1.8}_{-1.2}$ s  &$0.52^{+0.25}_{-0.17}$ s&$6.1^{+3.0}_{-2.0}$ s  &$26.7^{+13.1}_{-8.7}$ s    \\
  $^{288}$114&$10.09\pm0.07$& $0.8^{+0.32}_{-0.18}$ s  &$0.67^{+0.37}_{-0.27}$ s &$0.22^{+0.12}_{-0.08}$ s& $0.52^{+0.30}_{-0.19}$ s &$0.98^{+0.56}_{-0.40}$ s   \\
  $^{287}$114&$10.16\pm0.06$& $0.51^{+0.18}_{-0.10}$ s &$1.13^{+0.52}_{-0.40}$ s  &$0.16^{+0.08}_{-0.05}$ s& $1.79^{+0.85}_{-0.57}$ s &$7.24^{+3.43}_{-2.61}$ s  \\
  $^{286}$114&$10.35\pm0.06$& $0.16^{+0.07}_{-0.03}$ s &$0.14^{+0.06}_{-0.04}$ s &$0.05^{+0.02}_{-0.02}$ s & $0.11^{+0.05}_{-0.03}$ s &$0.19^{+0.08}_{-0.06}$ s \\
  $^{284}$113  &10.15 (6)& 0.48$^{+0.58}_{-0.17}$ s &1.55$^{+0.72}_{-0.48}$ s&0.43$^{+0.21}_{-0.13}$ s   &$2.4^{+1.2}_{-0.80}$ s    & 4.13$^{+1.94}_{-1.31}$ s \\
  $^{283}$113  & 10.26 (9)&100$^{+490}_{-45}$ ms    &201.6$^{+164.9}_{-84.7}$ms&222$^{+172}_{-96}$ ms   & $234^{+180}_{-100}$ ms & 937$^{+719}_{-402}$ ms   \\
   $^{285}$112&$9.29\pm0.06$& $34^{+17}_{-9}$ s &$75^{+41}_{-26}$ s &$13.22^{+7.25}_{-4.64}$ s&$127^{+69}_{-44}$ s  &$592^{+323}_{-207}$ s   \\
  $^{283}$112& $9.67\pm0.06$& $4.0^{+1.3}_{-0.7}$ s  &$5.9^{+2.9}_{-2.0}$ s &$0.95^{+0.48}_{-0.32}$ s&$9.6^{+4.9}_{-3.2}$ s  &$41.3^{+20.9}_{-13.8}$ s   \\
  $^{280}$111  & 9.87 (6)&  3.6 $^{+4.3}_{-1.3}$  s &1.9$^{+0.9}_{-0.6}$ s  &0.69$^{+0.33}_{-0.23}$ s   &$3.1^{+1.6}_{-1.05}$ s   & 5.70$^{+2.74}_{-1.84}$ s \\
  $^{279}$111  & 10.52(16)&170$^{+810}_{-80}$ ms    &9.6$^{+14.8}_{-5.7}$ ms&12.4$^{+19.9}_{-7.6}$ ms   &$10.9^{+17.8}_{-6.7}$ ms  & 45.3$^{+73.1}_{-27.6}$ms \\
  $^{279}$110&$9.84\pm0.06$& $0.18^{+0.05}_{-0.03}$ s &$0.40^{+0.18}_{-0.13}$ s &$0.08^{+0.04}_{-0.02}$ s& $0.65^{+0.31}_{-0.21}$ s &$2.92^{+1.4}_{-0.94}$ s  \\
  $^{276}$109  & 9.85 (6)& 0.72$^{+0.87}_{-0.25}$ s &0.45$^{+0.23}_{-0.14}$ s&0.19$^{+0.08}_{-0.06}$ s   & $0.65^{+0.33}_{-0.22}$ s  & 1.44$^{+0.68}_{-0.46}$ s \\
 $^{275}$109  & 10.48 (9)& 9.7$^{+46}_{-4.4}$  ms  &2.75$^{+1.85}_{-1.09}$ ms&4.0$^{+2.8}_{-1.6}$ ms    &$3.2^{+2.3}_{-1.3}$ ms  & 13.7$^{+9.6}_{-5.6}$ ms  \\
  $^{275}$108&$9.44\pm0.07$& $0.15^{+0.27}_{-0.06}$ s &$1.09^{+0.73}_{-0.40}$ s &$0.27^{+0.16}_{-0.10}$ s &$1.9^{+1.2}_{-0.72}$ s  &$8.98^{+5.49}_{-3.38}$ s  \\
  $^{272}$107  & 9.15 (6)&  9.8$^{+11.7}_{-3.5}$  s &10.1$^{+5.4}_{-3.4}$ s&5.12$^{+3.19}_{-1.58}$  s   & $17.6^{+10.2}_{-6.4}$ s  & 33.8$^{+17.9}_{-11.6}$ s  \\
  $^{271}$106&$8.65\pm0.08$& $2.4^{+4.3}_{-1.0}$ min &$1.0^{+0.8}_{-0.5}$ min &$0.33^{+0.28}_{-0.16}$ min&$1.8^{+1.5}_{-0.8}$ min  &$8.6^{+7.3}_{-3.9}$ min   \\
\end{tabular}
\end{ruledtabular}
\end{table*}

\begin{table*}[h]
\caption {\label{table3}Predicted $\alpha$-decay half-lives using
the GLDM, the formulae (1-4) and the VSS formulae. The $\alpha$ decay energies are
taken from the extrapolated data of Audi et al. \cite{Au03}.}
\begin{ruledtabular}
\begin{tabular}{lllllllllllllll}
$^{A}_{Z}$& Q &$T^{GLDM}_{1/2} $&$T^{form.}_{1/2} $&$T^{VSS}_{1/2} $&$^{A}_{Z}$& Q &$T^{GLDM}_{1/2} $&$T^{form.}_{1/2} $&$T^{VSS}_{1/2} $&$^{A}_{Z}$& Q &$T^{GLDM}_{1/2} $&$T^{form.}_{1/2} $&$T^{VSS}_{1/2} $\\
\hline
&&&\\
$^{293}_{118}$&12.30& 77 $\mu$s& 187 $\mu$s&592 $\mu$s &$^{292}_{117}$&11.60& 1.30 ms& 6.47 ms&13.33 ms &$^{291}_{117}$&11.90& 0.29  ms&0.32 ms &  1.23 ms \\ 
$^{291}_{115}$&10.00&  4.33 s &4.8 s & 21.9 s &$^{290}_{115}$&10.30& 0.62 s&4.2 s& 6.86 s &$^{289}_{116}$&11.70& 0.43 ms & 1.05 ms & 3.63 ms \\
$^{289}_{115}$&10.60& 97.4 ms& 113 ms &  482 ms  &$^{287}_{113}$& 9.34& 102 s & 99.4 s &  461 s   &$^{286}_{113}$& 9.68& 9.44 s& 61.5 s  & 92.5 s\\
$^{285}_{114}$&11.00& 5.1 ms&12 ms &  44.6 ms &$^{285}_{113}$&10.02& 0.99 s &1.0 s &  4.35 s  &$^{284}_{112}$& 9.30& 64.7 s  & 25.1 s &  47.3 s  \\
$^{283}_{111}$& 8.96& 6.01 min&5.5 min  & 25.73 min  &$^{282}_{112}$& 9.96& 0.772 s&0.297 s & 0.516 s&$^{282}_{111}$& 9.38& 18.6 s&99.8 s   & 158.4 s \\
$^{281}_{112}$&10.28& 0.102 s&0.2 s & 0.786 s  &$^{281}_{111}$& 9.64& 3.12  s&2.72 s & 11.96 s&$^{281}_{110}$& 8.96& 3.05 min&4.6 min   & 22.47 min \\
$^{280}_{112}$&10.62& 13.3 ms&25.4 ms & 8.62 ms  &$^{280}_{111}$& 9.98& 0.335 s&1.43 s & 2.79 s&$^{279}_{112}$&10.96& 2.06 ms &3.88 ms& 14.1 ms \\
$^{279}_{109}$& 8.70& 10.35 min&7.72 min & 36.32 min  &$^{278}_{112}$&11.38& 0.223 ms&0.083 ms & 0.121 ms&$^{278}_{111}$&10.72& 3.89 ms&12.5 ms  & 30.9 ms \\
$^{278}_{110}$&10.00& 148.5 ms&51.8 ms & 89.8 ms  &$^{278}_{109}$& 9.10&   31 s&143 s  & 240  s 
&$^{277}_{112}$&11.62& 0.069 ms&0.12 ms & 0.402 ms  \\
$^{277}_{111}$&11.18& 0.323 ms&0.28 ms& 1.073 ms  &$^{277}_{110}$&10.30& 23.1 ms&39 ms & 162 ms&$^{277}_{109}$& 9.50& 1.89  s&1.48 s  & 6.61 s   \\
$^{277}_{108}$& 8.40& 49.7 min&65.25 min  & 330.3 min  &$^{276}_{111}$&11.32& 0.157 ms&0.39 ms& 1.11 ms&$^{276}_{110}$&10.60& 4.03 ms&1.47 ms  & 2.35 ms  \\ 
$^{276}_{108}$& 8.80& 131 s&40.6 s& 75 s&$^{275}_{111}$&11.55&51.5 $\mu$s&42.3 $\mu$s&152 $\mu$s &$^{275}_{110}$&11.10& 0.26 ms &0.43 ms& 1.65 ms  \\ 
$^{274}_{111}$&11.60&41.4 $\mu$s&88.1 $\mu$s &258 $\mu$s &$^{274}_{110}$&11.40&55.5 $\mu$s&19.5 $\mu$s &28.7 $\mu$s&$^{274}_{109}$&10.50& 3.67 ms&9.84 ms  & 26.8 ms  \\ 
$^{274}_{108}$& 9.50& 0.92 s&0.3 s& 0.51 s&$^{274}_{107}$& 8.50& 9.94 min&48.45 min  & 70.98 min&$^{273}_{111}$&11.20& 0.33 ms&0.29 ms  & 0.96 ms \\ 
$^{273}_{110}$&11.37&0.067 ms&0.11 ms  & 0.39 ms  &$^{273}_{109}$&10.82& 0.61 ms&0.5 ms & 1.96 ms&$^{273}_{108}$& 9.90& 69.4 ms&101 ms & 441.6 ms \\
$^{273}_{107}$& 8.90& 28.8 s&21.1 s   & 92.8 s   &$^{272}_{110}$&10.76& 1.97 ms&0.697 ms  & 0.94 ms &$^{272}_{109}$&10.60& 2.34 ms&5.74 ms  & 15.02 ms \\ 
$^{272}_{108}$&10.10& 21.7 ms&6.9 ms & 10.9 ms&$^{272}_{106}$& 8.30& 24.9 min&6.38 min& 11.4 min   &$^{271}_{110}$&10.87& 1.12 ms & 1.79 ms&5.86 ms\\
$^{271}_{109}$&10.14& 37.5 ms&29.9 ms  & 105.6 ms&$^{271}_{108}$& 9.90& 79.2 ms&109.7 ms& 441.7 ms &$^{271}_{107}$& 9.50& 0.499 s&0.338 s & 1.40 s \\
$^{270}_{110}$&11.20& 0.199 ms&0.067 ms & 0.083 ms &$^{270}_{109}$&10.35& 10.7 ms&30 ms & 65 ms&
$^{270}_{108}$& 9.30& 4.48 s&1.4 s   & 2.02 s \\
$^{270}_{107}$& 9.30& 2.0 s&6.25 s    & 11.9 s   &$^{270}_{106}$& 9.10& 3.59 s&0.99 s   & 1.66 s& $^{270}_{105}$& 8.20& 24.38 min& 94.58 min&140.53 min  \\
$^{269}_{109}$&10.53& 3.75 ms&3.12 ms & 10.25 ms&$^{269}_{108}$& 9.63& 0.48 s&0.68 s   & 2.52 s  &$^{269}_{107}$& 8.84& 55.9 s&39 s   & 144.5 s  \\
$^{269}_{106}$& 8.80& 32.5 s&37.5 s  & 167.9 s&$^{269}_{105}$& 8.40& 4.96 min&3.01 min   & 12.93 min &$^{268}_{110}$&11.92& 6.3 $\mu$s& 1.84 $\mu$s& 2.1 $\mu$s \\ 
$^{268}_{109}$&10.73& 1.28 ms&3.07 ms & 7.15 ms&$^{268}_{108}$& 9.90& 85.7 ms&28.6 ms  & 37.7 ms &$^{268}_{107}$& 9.08& 9.86 s &35.4 s  & 55.5 s   \\ 
$^{268}_{106}$& 8.40& 12.1 min&3.4 min  & 5.1 min&$^{268}_{105}$& 8.20& 25.4 min&102.7 min& 140.5 min &$^{268}_{104}$& 8.10& 23.8 min&5.88 min & 10.2 min \\
$^{267}_{110}$&12.28&1.3 $\mu$s&1.57 $\mu$s& 4.4 $\mu$s&$^{267}_{109}$&10.87& 0.61 ms&0.49 ms  & 1.49 ms &$^{267}_{108}$&10.12& 22.1 ms&32.9 ms  & 112.5 ms \\ 
$^{267}_{107}$& 9.37& 1.33 s&0.97 s  & 3.36 s&$^{267}_{106}$& 8.64& 1.9 min&2.25 min & 9.3 min   &$^{267}_{105}$& 7.90& 330 min&205 min& 787 min  \\
$^{267}_{104}$& 7.80& 315  min&306 min  & 1494 min &$^{266}_{109}$&10.996& 0.32 ms&0.69 ms & 1.63 ms &$^{266}_{108}$&10.336& 6.26 ms&2.16 ms & 2.64 ms  \\
$^{266}_{107}$& 9.55& 0.41 s&1.21 s  & 2.21 s&$^{266}_{105}$& 8.19& 29.0 min&121.8 min& 152.5 min  &$^{266}_{104}$& 7.50& 81.47 h&20.09 h & 31.30 h\\
$^{265}_{109}$&11.07& 0.223 ms &0.178 ms& 0.498 ms&$^{265}_{107}$& 9.77&  99.7 ms&74.4 ms & 241 ms&$^{265}_{105}$& 8.49& 2.70 min&1.76 min   & 6.43 min  \\
$^{265}_{104}$& 7.78& 6.58 h&6.58 h  & 29.65 h&$^{264}_{107}$& 9.97& 29.9 ms&74.1 ms  & 151 ms   &$^{264}_{106}$& 9.21& 1.99 s  &0.60 s& 0.77 s \\
$^{264}_{105}$& 8.66& 46.1 s&154 s   & 232 s   &$^{264}_{104}$& 8.14& 19.2 min&5.03 min& 7.36 min   &$^{263}_{108}$&10.67& 1.03 ms &1.52 ms& 4.45 ms \\
$^{263}_{107}$&10.08& 15.5 ms&11.6 ms  & 34.9 ms &$^{263}_{105}$& 9.01& 3.65 s&2.4 s  & 8.27 s & $^{263}_{104}$& 8.49& 72.7 s &76.8 s  & 324.7 s \\
$^{262}_{107}$&10.30& 4.42 ms&9.51 ms  & 20.5 ms  &$^{262}_{106}$& 9.60&160.4 ms&47.5 ms & 56.7 ms&$^{262}_{105}$& 9.01& 4.06 s&10.9 s   & 18.2 s \\
$^{262}_{104}$& 8.49& 82.6 s&20.6 s   & 27.9 s   &$^{261}_{107}$&10.56& 1.04 ms&0.74 ms & 2.07 ms&$^{261}_{106}$& 9.80& 44.8 ms&56.1 ms  & 183.9 ms \\
$^{261}_{105}$& 9.22& 0.96 s&0.60 s   & 1.92 s   &$^{260}_{107}$&10.47& 1.77 ms&3.58 ms  & 7.62 ms &$^{260}_{104}$& 8.90& 4.09 s&1.08 s   & 1.35 s  \\
$^{259}_{106}$& 9.83& 39.4 ms&50.5 ms  & 152.3 ms &$^{259}_{105}$& 9.62& 69.0 ms&45.9 ms & 136.7 ms&$^{259}_{104}$& 9.12& 0.89 s& 0.93 s  & 3.38 s \\
$^{258}_{106}$& 9.67& 114 ms&36.1 ms   & 36 ms    &$^{258}_{105}$& 9.48& 0.18 s&0.42 s  & 0.74 ms &$^{258}_{104}$& 9.25& 380 ms&103 ms   & 120 ms  \\
$^{257}_{105}$& 9.23& 1.0 s&0.67 s    & 1.8  s   &$^{257}_{104}$& 9.04& 1.66 s&1.76 s  & 5.88 s&$^{256}_{105}$& 9.46& 230 ms&522 ms   & 848 ms  \\
$^{256}_{104}$& 8.93& 3.78 s&1.04 s   & 1.09 s   &$^{255}_{105}$& 9.72& 42.9 ms&28.9 ms & 72.4 ms&$^{255}_{104}$&9.058& 1.57 s&1.69 s   & 5.19 s  \\
$^{254}_{104}$& 9.38& 181 ms&51.9 ms   & 50.5 ms  &$^{253}_{104}$& 9.55& 63.1 ms&68.3 ms & 195.0 ms \\
\end{tabular}
\end{ruledtabular}
\end{table*}

The $\alpha$ decay process was described in 1928 \cite{Gam28,Con28} in terms of a quantum tunnelling through the potential barrier separating the mother nucleus energy and the total energy of the separated $\alpha$ particle and daughter nucleus.    
To describe the $\alpha$ emission two different approaches have been developed. The cluster-like theories suppose that the $\alpha$ particle is preformed in the nucleus with a certain preformation factor while the fission-like approaches consider that the $\alpha$ particle is formed progressively during the very asymmetric fission of the parent nucleus. The experimental investigation cannot unambiguously distinguish these two formation modes. However the possible one-body configurations play a minor role since in the quasi-molecular decay path investigated in the $\alpha$ decay process the potential barrier is governed by the balance between the repulsive Coulomb forces and the attractive proximity forces and the $Q_{\alpha}$ value;  consequently the barrier top is more external and lower than the pure Coulomb barrier and corresponds to two separated fragments. The difference between the two approaches appears mainly in the way the decay constant is determined. In the unified fission models \cite{Poen85,Roy00} the decay constant $\lambda$ is the product of the constant assault frequency $\nu_{0}$ and the barrier penetrability P while in the preformed cluster models \cite{Mal89,Lud92} a third factor is introduced : the cluster preformation probability $P_{0}$. 

Before the theoretical explanation and description of the $\alpha$ decay process, Geiger and Nuttal \cite{Gei11} observed a dependence of the $\alpha$ decay partial half-life T $^{expt}_{1/2,\alpha}$ on the mean $\alpha$ particle range for a fixed radioactive family and Geiger-Nuttal plots are now an expression of $log_{10}T_{\alpha}$ as a function of $ZQ^{-1/2}$. 
Since, different new relations have been proposed \cite{Vio66,Sobi89,Bro92,Poena96,Roy00,Po06} to calculate $log_{10}T_{\alpha}$ from the measured kinetic energy of the $\alpha$ particle via $E_{\alpha}=Q_{\alpha}A_d/(A_{\alpha}+A_d)$ or from $Q_{\alpha}$ given or extrapolated from mass formulae or tables. 

Recently, isotopes of the elements 112, 113, 114, 115, 116 and 118 have been synthetized in fusion-evaporation reactions using 
$^{209}$Bi, $^{233,238}$U, $^{242,244}$Pu, $^{243}$Am, $^{245,248}$Cm and $^{249}$Cf targets with $^{48}$Ca
and $^{70}$Zn beams and observed via their $\alpha$ decay cascades \cite{Ho00,Og99,Og04,Og05,Hof07,Og07,Mo07}. 
These recent experimental results have led to new theoretical studies on the $\alpha$ decay; for example within the relativistic mean field theory 
\cite{Sh05}, the DDM3Y interaction \cite{Ba04,Ch07}, the generalized liquid drop model (GLDM) \cite{Zh06,Zh07} and Skyrme-Hartree-Fock mean-field model \cite{Pe07}. The predicted half-lives against $\alpha$ decay of these transuranium nuclei obtained with a semiempirical formula taking into account the magic numbers have also been compared with the analytical superasymmetric fission model results and the universal curves and the experimental data \cite{Po06}. 

In previous studies \cite{Roy00,Mou01} both theoretical description and analytical formulas were presented for the $\alpha$ emission. Within a generalized liquid drop model including the proximity effects between the $\alpha$ particle and the daughter nucleus and adjusted to reproduced the experimental Q value the $\alpha$ emission half-lives were deduced from the WKB barrier penetration probability as for a spontaneous asymmetric fission. The RMS deviation between the theoretical and experimental values of $log_{10}T_{\alpha}$ was 0.63 for a data set of 373 emitters having an $\alpha$ branching
ratio close to one and 0.35 for the subset of 131 even-even nuclides. A fitting procedure led to the following empirical formulas respectively for the 131 even(Z)-even(N), 106 even-odd, 86 odd-even and 50 odd-odd nuclei
. A and Z are the mass and charge numbers of the mother nucleus. The rms deviation are respectively 0.285, 0.39, 0.36 and 0.35.
\begin{equation}
log_{10}\left \lbrack T_{1/2}(s)\right \rbrack=-25.31
-1.1629A^{1/6}\sqrt{Z}+\frac{1.5864Z}{\sqrt{Q_{\alpha}}},
\end{equation}  
\begin{equation}
log_{10}\left \lbrack T_{1/2}(s)\right \rbrack=-26.65
-1.0859A^{1/6}\sqrt{Z}+\frac{1.5848Z}{\sqrt{Q_{\alpha}}},
\end{equation}  
\begin{equation}
log_{10}\left \lbrack T_{1/2}(s)\right \rbrack=-25.68
-1.1423A^{1/6}\sqrt{Z}+\frac{1.592Z}{\sqrt{Q_{\alpha}}},
\end{equation}  
\begin{equation}
log_{10}\left \lbrack T_{1/2}(s)\right \rbrack=-29.48
-1.113A^{1/6}\sqrt{Z}+\frac{1.6971Z}{\sqrt{Q_{\alpha}}}.
\end{equation}  
 
Since new $\alpha$ decays have been observed and their 
partial $\alpha$ decay half-lives T$^{expt}_{1/2,\alpha}$ have been measured \cite{Og04,Og05,Hof07,Og07,Mo07,Pe07,An99,Gar00,Ma00,Ke01,Se06,Lep07}. They are compared in the Tables 1 and 2 with the calculated values from the above-mentioned formulae using the measured $Q_{\alpha}$ values. The table 2 displays also the results obtained with the DDM3Y effective
interaction \cite{Ba04,Ch07}, the GLDM \cite{Roy00,Zh06} and the Viola-Seaborg formulae with Sobiczewski constants \cite{Vio66,Sobi89}.

A quite good agreement appears in the table 1 in the whole mass range confirming the accuracy of the formulas (1-4) and their usefulness for new predictions. The table 2 focuses on the heaviest elements for which the uncertainties both on the experimental Q value and $\alpha$ decay half-lives are larger since only some $\alpha$ cascades have been observed.  
The results obtained with the DDM3Y effective interaction agree with the
experimental data as the ones calculated from the GLDM and largely better than
the VSS calculations which give systematically longer half-lives. This shows that a GLDM taking account the
proximity effects, the mass asymmetry and quasi-molecular shapes is sufficient to reproduce the $\alpha$ decay potential
barriers when the experimental $Q_{\alpha}$ value is known and proves that the double folding potential
obtained using M3Y effective interaction supplemented
by a zero-range potential for the single-nucleon exchange is also very
appropriate to describe the $\alpha$ decay process. The DDM3Y results are on an average slightly larger than the experimental data while the GLDM values are slightly
lower than the measured values. The values obtained using the formulae (1-4) and, then, only A, Z and $Q_{\alpha}$ are close to the values derived from the DDM3Y interaction and in agreement with the still rough experimental data.   
The fact that the partial $\alpha$ decay half-lives of these superheavy elements follow these simple formulae 
seems to prove that the experimental data are consistent with the formation of a cold and relatively compact composite nuclear system. The shell effects are implicitly contained in the $Q_{\alpha}$ value but difficult to disentangle. 

Thus predictions of the partial $\alpha$ decay half-lives of still unknown superheavy nuclei within the formulas (1-4) seem reliable and are displayed in the table 3. The values obtained using the GLDM and the VSS expressions are also given for comparison. The assumed $\alpha$ decay energies are calculated from the atomic mass evaluation of Audi et al. \cite{Au03} since the agreement with the experimental data on the mass of the known heaviest elements is very satisfactory. It may be useful for future experimental assignment and identification. 

In conclusion, formulae already presented to determine the partial $\alpha$ decay half-lives have been checked on new experimental data in the whole mass range and the correct agreement allows to provide predictions for the  partial $\alpha$ decay half-lives of still unknown superheavy nuclei.


\end{document}